\documentclass[carbon,article,accept,pdftex,moreauthors]{Definitions/mdpi}
\UseRawInputEncoding
\usepackage{amsmath}
\usepackage{amsfonts}
\usepackage{amssymb,bm}
\usepackage{siunitx}
\usepackage{color}
\usepackage{braket}
\usepackage{graphics}

\pubvolume{9}
\issuenum{3}
\articlenumber{64}
\pubyear{2023}
\copyrightyear{2023}
\externaleditor{Academic Editor: Matteo Strozzi and Ahmet Sinan Oktem}
\datereceived{30 May 2023 }
\daterevised{16 June 2023 } 
\dateaccepted{25 June 2023 }
\datepublished{4 July 2023 }
\hreflink{https://doi.org/10.3390/c9030064}
\pdfoutput=1

\Title{Casimir-Polder Force on Atoms or Nanoparticles from the Gapped and Doped Graphene:
Asymptotic Behavior at Large Separations}

\TitleCitation{Casimir-Polder Force on Atoms or Nanoparticles from the Gapped and Doped Graphene:
Asymptotic Behavior at Large Separations}

\Author{{Galina L. Klimchitskaya} ${}^{1,2}$\orcidA{}
and
Vladimir M. Mostepanenko ${}^{1,2,3}$*\orcidB{}}

\AuthorNames{Galina L. Klimchitskaya,
Vladimir M. Mostepanenko}

\AuthorCitation{Klimchitskaya, G.L.;  Mostepanenko, V.M.}

\address{%
$^{1}$ \quad Central Astronomical Observatory at Pulkovo of the Russian Academy of Sciences, 196140 Saint Petersburg, Russia\\
$^{2}$ \quad Peter the Great Saint Petersburg
Polytechnic University, 195251 Saint Petersburg, Russia\\
$^{3}$ \quad  {Kazan Federal University}, 420008 Kazan, Russia}

\corres{Correspondence: vmostepa@gmail.com}

\abstract{The Casimir-Polder force acting on atoms and nanoparticles spaced at large
separations from real
graphene sheet possessing some energy gap and chemical potential is investigated in the
framework of the Lifshitz theory. The reflection coefficients expressed via the polarization
tensor of graphene found based on the first principles of thermal quantum field theory are used.
It is shown that for graphene the separation distances starting from which the zero-frequency
term of the Lifshitz formula contributes more than 99\% of the total Casimir-Polder force are
less than the standard thermal length. According to our results, however, the classical limit
for graphene, where the force becomes independent on the Planck constant, may be reached at
much larger separations than the limit of large separations determined by the zero-frequency
term of the Lifshitz formula depending on the values of the energy gap and chemical potential.
The analytic asymptotic expressions for the zero-frequency term of the Lifshitz formula at
large separations are derived. These asymptotic expressions agree up to 1\% with the results
of numerical computations starting from some separation distance which increases with
increasing energy gap and decreases with increasing chemical potential. Possible applications
of the obtained results are discussed. }


\begin{document}

\section{Introduction}

The Casimir-Polder force \cite{1} acts between electrically neutral small bodies (atoms,
nanoparticles) and material surfaces. This force is induced by the zero-point and thermal
fluctuations of the electromagnetic field which have their origin in the microscopic charges
and currents occurring inside all material bodies. It is a generalization of the van der Waals
force to separation distances where the relativistic effects already make a pronounced
impact on the force value. This typically happens at separations exceeding several nanometers.

The unified theory of the atom-plate van der Waals and Casimir-Polder forces was
developed by Lifshitz \cite{2,3,4}. Given the dynamic polarizability $\alpha(\omega)$ of an
atom or a nanoparticle and the dielectric function of a material plate, one can calculate the
Casimir-Polder force in the framework of the Lifshitz theory. Calculations of this kind have
been performed for different atoms, nanoparticles, and plate materials
\cite{5,6,7,8,9,10,11,12,13,14,15,16,17,18,19,20}. The obtained results were found to
be important in explaining the crucially new physical phenomena of quantum reflection
\cite{21,22, 23,24,25,26,27,28,29,30} and Bose-Einstein condensation \cite{31,32,33,34,35}.
It should be noted that the original Lifshitz theory was formulated for the physical systems
which are in the state of thermal equilibrium with the environment. The generalization of
this theory for situations out of thermal equilibrium (for instance, when the plate temperature
is different from that of the environment) was performed in \cite{36,37,38,39,40,41}. It
was applied to calculate the nonequilibrium Casimir-Polder force in different cases
including the experimental configurations \cite{42,43,44,45}.

Considerable recent attention has been focussed on graphene which is a two-dimensional
sheet of carbon atoms packed in the hexagonal lattice \cite{46}. Currently graphene finds
expanding applications in both fundamental physics and in nanotechnology. Specifically,
the Casimir-Polder force acting on atoms \cite{47,48,49,50,51,52,53,54,55,56,57,58,59}
and nanoparticles \cite{60,61,62,63,64,65} from graphene and graphene-coated
substrates has become the topic of a large body of research. Graphene was demonstrated
to have properties described by the Dirac model \cite{46,66,67}, i.e., at energies below
approximately 3 eV \cite{68} electrons in graphene can be considered as a set of massless or
light quasiparticles governed by the Dirac equation in two spatial dimensions, where the speed
of light $c$ is replaced by the Fermi velocity $v_F\approx c/300$. This made it possible
to find the polarization tensor of graphene at any temperature \cite{69,70,71,72}, which
is equivalent to the spatially nonlocal dielectric functions, the transverse one and the
longitudinal one. These results were used in \cite{48,49,50,51,53,54,55,56,57,58,59,65}
to calculate the Casimir-Polder force acting on an atom or a nanoparticle from graphene
in the framework of the Lifshitz theory.

An important question is how soon the Casimir-Polder force from graphene approaches its
limiting form given by the zero-frequency term in the Lifshitz formula, which is reached at large
separations (high temperatures). In \cite{73}, the asymptotic behavior of the Casimir-Polder
interaction was investigated in the case of an undoped graphene sheet possessing
zero chemical potential. However, real graphene sheets are characterized not only by
the energy gap in the spectrum of quasiparticles $\Delta=2mv_F^2$, where $m$ is small
but nonzero mass of quasiparticles \cite{46,74,75}, but they are also doped, i.e., their crystal
lattice contains some fraction of foreign atoms. This can be described by a nonzero value of
the chemical potential $\mu$ depending on the doping concentration \cite{76}.

In this article, we examine the behavior of the Casimir-Polder force between atoms
(nanoparticles) and real graphene sheets in the limit of large separations (high temperatures)
as a function of the atom-plate separation $a$, the energy gap $\Delta$, and the chemical
potential $\mu$. First, we demonstrate that the term of the Lifshitz formula at zero
Matsubara frequency contributes more than 99\% of the force magnitude at separations
exceeding some value $a_0$, which is distinctly less than the standard thermal length
$\hbar c/k_BT)\approx 7.6\mu$m at room temperature $T=300$K (here $k_B$ is the
Boltzmann constant). The value of $a_0$ decreases with increasing $\Delta$. For
sufficiently small $\Delta$, $a_0$ increases with increasing $\mu$, but for a larger
$\Delta$ the dependence of $a_0$ on $\mu$ becomes nonmonotonic.

Then we compare the large-separation Casimir-Polder force from graphene, given by the
zero-frequency term of the Lifshitz formula, with that from an ideal metal plane. It is
shown that for a fixed energy gap an agreement between these two quantities becomes better
with increasing chemical potential of a graphene sheet.

Next, we derive simple asymptotic expressions for the zero-frequency contribution to the
Lifshitz formula at large separations and find how it agrees with the results of numerical
computations. For this purpose, we use the zero-frequency term of the Lifshitz formula
with reflection coefficients expressed via the polarization tensor of graphene. The polarization
tensor is calculated using several small parameters. The
analytic asymptotic expressions for the large-separation Casimir-Polder force are derived
for any values of the energy gap and chemical potential of a graphene sheet.

The derived asymptotic expressions are compared with numerical computations of the
Casimir-Polder force at large separations. The application region of the analytic
asymptotic results is determined. We show that with increasing energy gap an agreement
between the asymptotic and computational results becomes worse, whereas, at the
same separation, an increase of the chemical potential brings the asymptotic results in better
agreement with the results of numerical computations.

The article is organized as follows. In Section 2, we present the Lifshitz formula for the
Casimir-Polder force and reflection coefficients for the case of gapped and doped graphene
in terms of the polarization tensor. Section 3 contains the exact expression and numerical
computations of the Casimir-Polder force at large separations. In Section 4, the analytic
asymptotic expressions for the Casimir-Polder force are derived. In Section 5, the
asymptotic results for the Casimir-Polder force are compared with the results of numerical
computations. Section 6 contains a discussion and Section 7 --- our conclusions.

\section{The Lifshitz Formula and Reflection Coefficients for Gapped and Doped
Graphene}
\newcommand{\Mr}{{r_{\rm TM}(i\zeta_l,y)}}
\newcommand{\Er}{{r_{\rm TE}(i\zeta_l,y)}}
\newcommand{\Fv}{{\mbox{$\tilde{v}_F$}}}
\newcommand{\Sv}{{\mbox{$\tilde{v}_F^2$}}}
\newcommand{\tP}{{\mbox{$\widetilde{\Pi}$}}}

The Casimir-Polder force between an atom or a nanoparticle and any plane surface is
expressed by the following Lifshitz formula, which we present in terms of dimensionless
variables \cite{54,77}

\begin{equation}
F(a,T)=-\frac{k_BT}{8a^4}\sum_{l=0}^{\infty}{\vphantom{\sum}}^{\prime}\alpha_l
\int\limits_{\zeta_l}^{\infty}y\,dy\,e^{-y}\left[(2y^2-\zeta_l^2)\Mr-\zeta_l^2\Er\right].
\label{eq1}
\end{equation}

\noindent
Here, the prime on the sum in $l$  means that the term with $l=0$ is divided by 2,
$\zeta_l=\xi_l/\omega_c$, where $\xi_l=2\pi k_BTl/\hbar$ ($l=0,\,1,\,2,\,\ldots$)
are the Matsubara frequencies, $\omega_c=c/(2a)$ is the characteristic frequency,
and $\alpha_l=\alpha(i\zeta_l\omega_c)$. The dimensionless variable $y$ is defined as

\begin{equation}
y=2aq_l=2a\left(k_{\bot}^2+\frac{\xi_l^2}{c^2}\right)^{1/2},
\label{eq2}
\end{equation}

\noindent
where $k_{\bot}$ is the magnitude of the wave vector projection on the plane of
graphene, and $r_{\rm TM,TE}$ are the reflection coefficients on graphene for the
transverse magnetic (p) and transverse electric (s) polarizations of the electromagnetic
field. Note that both the dynamic polarizability $\alpha_l$ and the reflection
coefficients $r_{\rm TM,TE}$ are calculated at the pure imaginary frequencies $i\zeta_l$.

The reflection coefficients in (\ref{eq1}) are expressed via the dimensionless
polarization tensor of graphene \cite{54}

\begin{equation}
\Mr=\frac{y\tP_{00,l}}{y\tP_{00,l}+2(y^2-\zeta_l^2)},
\qquad
 \Er=-\frac{\tP_{l}}{\tP_{l}+2y(y^2-\zeta_l^2)},
\label{eq3}
\end{equation}

\noindent
where the components of the dimensionless $\tP_{\beta\gamma}$ and dimensional
$\Pi_{\beta\gamma}$ tensors ($\beta,\gamma=0,\,1,\,2$) are connected by

\begin{equation}
\tP_{\beta\gamma,l}\equiv\tP_{\beta\gamma}(i\zeta_l.y)=
\frac{2a}{\hbar}\Pi_{\beta\gamma}(i\xi_l,k_{\bot}).
\label{eq4}
\end{equation}

The dimensionless quantity $\tP_l$ in (\ref{eq2}) is defined as

\begin{equation}
\tP_{l}\equiv
\frac{(2a)^3}{\hbar}(k_{\bot}^2\Pi_{\beta,l}^{\,\beta}-q_l^2\Pi_{00.l})=
(y^2-\zeta^2)\tP_{\beta,l}^{\,\beta}-y^2\tP_{00.l}
\label{eq5}
\end{equation}

\noindent
with a summation over $\beta$. The arguments of the polarization tensor components
are omitted for brevity.

As mentioned in Introduction, the polarization tensor of graphene is equivalent to
the spatially nonlocal transverse and longitudinal dielectric functions defined in
two-dimensional space \cite{78,79}. Using the dimensionless variables, one obtains

\begin{equation}
\varepsilon ^{\rm L}(i\zeta_l,y)=1+\frac{1}{2\sqrt{y^2-\zeta_l^2}}\,\tP_{00,l},
\qquad
\varepsilon ^{\rm Tr}(i\zeta_l,y)=1+\frac{1}{2\zeta_l^2\sqrt{y^2-\zeta_l^2}}\,\tP_{l}.
\label{eq6}
\end{equation}

The explicit expression for $\tP_{00,l}$ in terms of the dimensionless variables
$\zeta_l$ and $y$ is presented in \cite{54}. After identical transformations it can be put
in a more convenient form

\begin{eqnarray}
&&
\tP_{00,l}=\alpha\frac{y^2-\zeta_l^2}{p_l}\Psi(D_l)+\frac{16\alpha ak_BT}{\Sv\hbar c}
\ln\left[\left(e^{-\frac{\Delta}{2k_BT}}+e^{\frac{\mu}{k_BT}}\right)
\left(e^{-\frac{\Delta}{2k_BT}}+e^{-\frac{\mu}{k_BT}}\right)\right]
\nonumber\\
&&\label{eq7}\\[-2mm]
&&-\frac{4\alpha p_l}{\Sv}\int\limits_{D_l}^{\infty}du \,w_l(u,y)\,{\rm Re}
\frac{p_l-p_lu^2+2i\zeta_l u}{[p_l^2-p_l^2u^2+\Sv(y^2-\zeta_l^2)D_l^2+
2i\zeta_lp_lu]^{1/2}}.
\nonumber
\end{eqnarray}

\noindent
Here, $\alpha=e^2/(\hbar c)$ is the fine structure constant,

\begin{equation}
\Psi(x)=2\left[x+(1-x^2){\rm arctan}\frac{1}{x}\right],
\qquad
p_l=\sqrt{\Sv y^2+(1-\Sv)\zeta_l^2},
\label{eq8}
\end{equation}

\noindent
$\Fv=v_F/c$ is the dimensionless Fermi velocity,

\begin{equation}
w_l(u,y)=\frac{1}{e^{B_lu+\frac{\mu}{k_BT}}+1}+
\frac{1}{e^{B_lu-\frac{\mu}{k_BT}}+1}
\label{eq9}
\end{equation}

\noindent
and, finally,

\begin{equation}
D_l\equiv D_l(y)=\frac{2a\Delta}{\hbar cp_l}, \qquad
B_l\equiv B_l(y)=\frac{\hbar cp_l}{4ak_BT}.
\label{eq10}
\end{equation}

In a similar way, the combination of the components of the polarization tensor $\tP_l$
entering Equation (\ref{eq3}) can be written in the form \cite{54}

\begin{eqnarray}
&&
\tP_{l}=\alpha(y^2-\zeta_l^2)p_l\Psi(D_l)-\frac{16\alpha ak_BT\zeta_l^2}{\Sv\hbar c}
\ln\left[\left(e^{-\frac{\Delta}{2k_BT}}+e^{\frac{\mu}{k_BT}}\right)
\left(e^{-\frac{\Delta}{2k_BT}}+e^{-\frac{\mu}{k_BT}}\right)\right]
\nonumber\\
&&\label{eq11}\\[-2mm]
&&+\frac{4\alpha p_l^2}{\Sv}\int\limits_{D_l}^{\infty}du \,w_l(u,y)\,{\rm Re}
\frac{\zeta_l^2-p_l^2u^2+\Sv(y^2-\zeta_l^2)D_l^2+
2i\zeta_l p_l u}{[p_l^2-p_l^2u^2+\Sv(y^2-\zeta_l^2)D_l^2+
2i\zeta_lp_lu]^{1/2}}.
\nonumber
\end{eqnarray}

By using Equations (\ref{eq1}), (\ref{eq3}), (\ref{eq7}), and(\ref{eq11}) one can compute
the Casimir-Polder force between an atom or a nanoparticle and a real graphene sheet
characterized by some energy gap and chemical potential.

\section{The Casimir-Polder Force at Large Separations}

It is well known that at large separations or, equivalently, at high temperatures the
dominant contribution to the Casimir-Polder force is given by the term of (\ref{eq1})
with $l=0$ \cite{77,80}. For atoms and nanoparticles interacting with the three-dimensional
plates made of ordinary materials the zero-frequency term of the Lifshitz formula is
approximately equal to the total force already at the thermal length
$\hbar c/(k_BT)\approx 7.6~\mu$m at room temperature. Below we demonstrate that for
graphene the zero-frequency term determines the total force value at even smaller
separations depending on the energy gap $\Delta$ and chemical potential $\mu$.

The zero-frequency term of the Lifshitz formula is obtained by separating the component
with $l=0$ from (\ref{eq1})

\begin{equation}
F_0(a,T)=-\frac{k_BT}{8a^4}\alpha_0\int\limits_{0}^{\infty}y^3dy\,e^{-y}r_{\rm TM}(0,y),
\label{eq12}
\end{equation}

\noindent
where the reflection coefficient from (\ref{eq3}) simplifies to

\begin{equation}
r_{\rm TM}(0,y)=\frac{\tP_{00,0}(y)}{\tP_{00,0}(y)+2y}.
\label{eq13}
\end{equation}

\noindent
Note that the TE reflection coefficient does not contribute to (\ref{eq12}) because
in (\ref{eq1}) taken at $l=0$ it is multiplied by zero.
Here we have explicitly indicated the argument $y$ of the polarization tensor.
The component of the dimensionless polarization tensor $\tP_{00,0}$ is obtained
from (\ref{eq7}) by putting $\zeta_0=0$

\begin{eqnarray}
&&
\tP_{00,0}(y)=\frac{\alpha y}{\Fv}\Psi(D_0)+\frac{16\alpha ak_BT}{\Sv\hbar c}
\ln\left[\left(e^{-\frac{\Delta}{2k_BT}}+e^{\frac{\mu}{k_BT}}\right)
\left(e^{-\frac{\Delta}{2k_BT}}+e^{-\frac{\mu}{k_BT}}\right)\right]
\nonumber\\[0mm]
&&\label{eq14}\\[-2mm]
&&-\frac{4\alpha y}{\Fv}\int\limits_{D_0}^{\sqrt{1+D_0^2}}du \,w_0(u,y)\,
\frac{1-u^2}{\sqrt{1-u^2+D_0^2}},
\nonumber
\end{eqnarray}

\noindent
where

\begin{equation}
D_0=\frac{2a\Delta}{\hbar c\Fv y}\, , \qquad
B_0=\frac{\hbar c\Fv y}{4ak_BT}
\label{eq15}
\end{equation}

\noindent
and $w_0(u,y)$ is defined in (\ref{eq9}) with $l=0$ and $B_0$ from (\ref{eq15}).

Now we perform numerical computations in order to find such separation $a_0$ that  at all
separations $a\geqslant a_0$ the quantity $F_0$ from (\ref{eq12}) contributes no less
than 99\% of the total Casimir-Polder force (\ref{eq1}). This is done for heavy atoms,
for instance, Rb and for nanoparticles.
It is apparent that the value of $a_0$ depends on the energy gap and chemical potential
of the specific graphene sheet,  so that $a_0=a_0(\Delta,\mu)$. For this purpose, first
we compute $F_0$ from (\ref{eq12}) as a function of separation using Equations
(\ref{eq13})--(\ref{eq15}). All computations here and below are performed at room
temperature, $T=300~$K, in the range of $\Delta$ from 0.001~eV to 0.2~eV with a step
of 0.01~eV for 4 values of $\mu=0$, 25, 75, and 150~meV.
Similar computations of the total Casimir-Polder force $F$ from (\ref{eq1}) were
performed by Equations (\ref{eq1}), (\ref{eq3}), (\ref{eq7}), and (\ref{eq11}) at
separations exceeding $1~\mu$m where, without the loss of accuracy, one can use an
approximation of the static atomic polarizability $\alpha_l\approx\alpha(0)=\alpha_0$
\cite{77}. The point is that computations of the Casimir-Polder force
from graphene by (\ref{eq1}) using the frequency-dependent polarizability
$\alpha(\omega)$ show \cite{54} that even for light atoms like He$^*$ the
value of $a_0$ is above 1 $\mu$m. In so doing, for heavy atoms like Rb
and for nanoparticles all $\alpha_l$ giving contribution to the result
(i.e., with $l\leq 6$ at $a=1 \mu$m, $l\leq 3$ at $a=2 \mu$m, and
$l\leq 2$ at $a=3 \mu$m) are approximately equal to $\alpha_0$.

\begin{figure}[H]
\vspace*{-7.2cm}
\centerline{\hspace*{-2.7cm}
\includegraphics[width=4.5in]{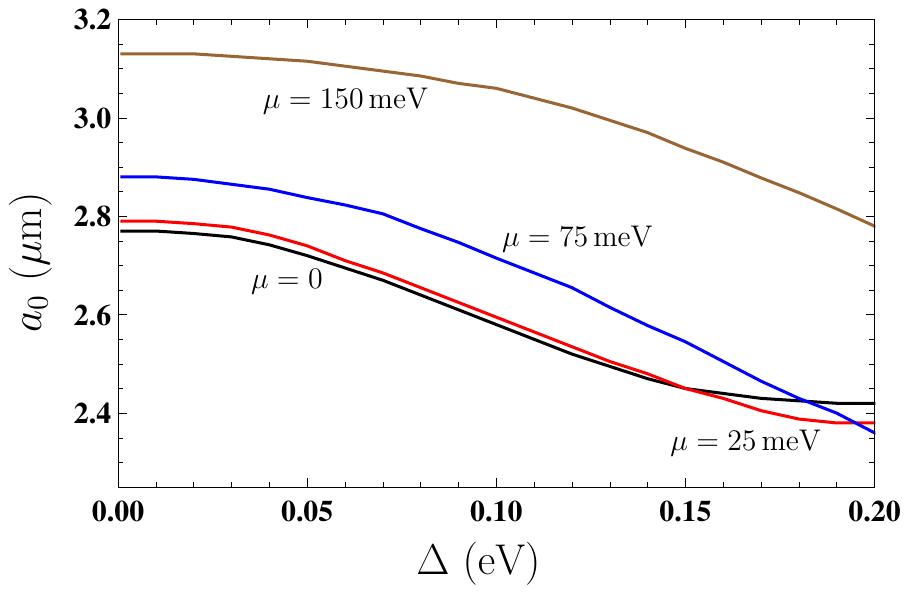}}
\vspace*{-3.7cm}
\caption{\label{fg1}
Minimum separation between an atom (nanoparticle) and real graphene sheet for which the
zero-frequency term of the Lifshitz formula contributes no less than 99\% of the total
Casimir-Polder force at $T=300~$K is shown as a function of the energy gap by the four
lines for the chemical potential equal to 0, 25, 75, and 150~meV. }
\end{figure}

Using the obtained computational results, in Figure~\ref{fg1} we plot $a_0$ as a
function of the energy gap of graphene $\Delta$ by the four lines (black, red, blue,
and brown) for the chemical potential $\mu$  equal to 0, 25, 75, and 150~meV, respectively.
{}From Figure~\ref{fg1} it is seen that the value of $a_0$ decreases with increasing
energy gap. For $\Delta<0.15~$eV the value of $a_0$ increases with increasing $\mu$,
but for larger $\Delta$ this already not the case. Specifically, the value of $a_0$
for a graphene sheet with $\mu=0$ may become larger than for sheets with $\mu=25$ and
75~meV. Intuitively it is clear that increasing $\mu$
brings graphene closer to an ideal metal and, thus, leads to an
increase of $a_0$. To the contrary, an increase of $\Delta$ results in
decreasing $a_0$ due to the suppressed impact of the thermal effects
on the Matsubara terms in (\ref{eq1}) with $l\geq 1$. The actual value
of $a_0$ for small $\mu$ and large $\Delta$ results from the interplay
between these two effects.
By and large, the value of $a_0$ for gapped and doped graphene is distinctly
less than the thermal length for ordinary materials equal to $7.6~\mu$m.

It is interesting also to compare the Casimir-Polder force from gapped and doped graphene
at large separations $F_0$ with that from an ideal metal plane given by \cite{77}

\begin{equation}
F_0^{\,\rm IM}(a,T)=-\frac{3k_BT}{4a^4}\alpha_0.
\label{eq16}
\end{equation}

\noindent
This is so-called classical limit because the force does not depend on the Planck
constant.

To understand where the large-separation Casimir-Polder forces from gpaphene and
from an ideal metal plane come together, we compute the relative quantity
\begin{equation}
\delta F_0(a,T)=\frac{F_0(a,T)-F_0^{\,\rm IM}(a,T)}{F_0^{\,\rm IM}(a,T)}.
\label{eq16}
\end{equation}

\begin{figure}[H]
\vspace*{-5.2cm}
\centerline{\hspace*{-2.7cm}
\includegraphics[width=5.in]{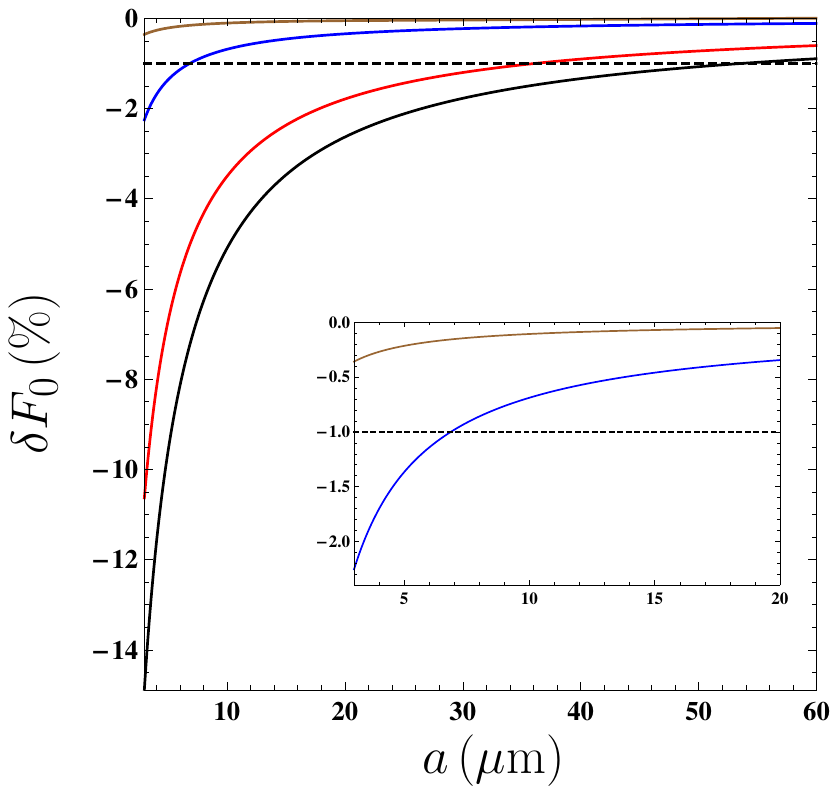}}
\vspace*{-4.7cm}
\caption{\label{fg2}
The relative deviation between the high-separation behaviors of the Casimir-Polder
force from graphene sheet with $\Delta=0.2~$eV and an ideal metal plane at $T=300~$K
is shown as a function of the energy gap by the four lines  counted from bottom to top
for the chemical potential equal to 0, 25, 75, and 150~meV.
In the inset, the two lines for $\mu=75$ and 150~meV (bottom and top, respectively)
are shown at short separations on an enlarged scale. The dashed lines indicate the
border of the 1-percent relative deviation. }
\end{figure}

The computational results for a graphene sheet with $\Delta=0.2~$eV are shown in
Figure~\ref{fg2} at $T=300~$K as a function of separation
by the four lines (black, red, blue, and brown) from bottom to top for the
chemical potential $\mu$  equal to 0, 25, 75, and 150~meV, respectively.
In the inset, the behavior of blue and brown lines at short separations
($\mu=75$ and 150~eV) is shown on an enlarged scale with better resolution.
The dashed lines indicate the border of the 1-percent relative deviation
between the large-separation behavior of the Casimir-Polder forces from a
graphene sheet and an ideal metal plane. The separation region $a\geqslant 3~\mu$m
is considered where, according to Figure~\ref{fg1}, $F_0$  represents
the large-separation behavior of the Casimir-Polder force.

As is seen in Figure~\ref{fg2}, for $\Delta=0.2~$eV, $\mu=150~$meV (the top line) the
Casimir-Polder forces from graphene and from an ideal metal plane agree within 1\% at
all separations considered. With decreasing $\mu$ to 75, 25, and 0~meV the agreement within
1\% occurs at separations exceeding 7, 36, and $54~\mu$m,  respectively.
This result is physically natural if to take into account that larger $\mu$ correspond
to larger doping concentration, i.e., make graphene more akin to an ideal metal plane.
Thus, for graphene sheets with relatively low chemical potential the classical limit
is reached only at rather large separation distances.

In the next section, we obtain simple asymptotic expressions for the quantity $F_0$
from (\ref{eq12}), which allow to calculate the Casimir-Polder force from gapped
and doped graphene at large separations do not using complicated expressions for
the polarization tensor.

\section{Asymptotic Expressions for the Casimir-Polder Force}
\newcommand{\pP}{{\mbox{$\widetilde{\Pi}_{00,0}(1)$}}}

We consider the Casimir-Polder force (\ref{eq12}) where the reflection coefficient
$r_{\rm TM}$ is given by (\ref{eq13}) and the polarization tensor is expressed by
(\ref{eq14}) with the notations (\ref{eq15}). We seek for the asymptotic expression of
(\ref{eq14}) and (\ref{eq12}) under the condition

\begin{equation}
\frac{2ak_BT}{\Fv\hbar c}=\frac{k_BT}{\Fv\hbar \omega_c}\gg 1.
\label{eq18}
\end{equation}

\noindent
At $T=300~$K this condition is well satisfied for $a>0.2~\mu$m, i.e., is not restrictive.

The reflection coefficient (\ref{eq13}) can be identically rewritten in the form

\begin{equation}
r_{\rm TM}(0,y)=1-\frac{2y}{\tP_{00,0}(y)+2y}.
\label{eq19}
\end{equation}

As is seen from (\ref{eq14}), the parameter (\ref{eq18}) stands in front of the second
contribution to $\tP_{00,0}$ by making it much larger than unity. Note that this contribution
does not depend on $y$. Simultaneously, the main contribution to (\ref{eq12}) is given by
$y\sim 1$. Because of this, one can replace $y$ with unity in the denominator of (\ref{eq19})
and neglect by 2 in comparison with $\tP_{00,0}$. As a result, (\ref{eq19}) takes the form

\begin{equation}
r_{\rm TM}(0,y)\approx 1-\frac{2y}{\pP}.
\label{eq20}
\end{equation}

Substituting (\ref{eq20}) in (\ref{eq12}) and integrating with respect to $y$, one obtains
the asymptotic expression

\begin{equation}
F_0^{as}(a,T)\approx -\frac{3k_BT\alpha_0}{4a^4}\left[1-\frac{8}{\pP}\right]=
F_0^{\,\rm IM}(a,T)\left[1-\frac{8}{\pP}\right],
\label{eq21}
\end{equation}

\noindent
where $F_0^{\,\rm IM}$ is the Casimir-Polder force from an ideal metal plane at large separations
defined in (\ref{eq16}). Note that $F_0^{as}(a,T)$ depends on the Planck constant $\hbar$ through
the polarization tensor of graphene $\pP$.

Now we deal with the asymptotic expressions for the polarization tensor $\pP$ and start from
the case $\Delta=0$, $\mu\neq 0$. In this case, we have from ({\ref{eq15}) $D_0=0$ and
from (\ref{eq8}) $\Psi(0)=\pi$. Because of this (\ref{eq14}) simplifies to

\begin{eqnarray}
&&
\pP\approx\frac{\alpha \pi}{\Fv}+\frac{8\alpha k_BT}{\Sv\hbar \omega_c}
\ln\left[\left(1+e^{\frac{\mu}{k_BT}}\right)
\left(1+e^{-\frac{\mu}{k_BT}}\right)\right]
\nonumber\\[0mm]
&&\label{eq22}\\[-2mm]
&&-\frac{4\alpha}{\Fv}\int\limits_{0}^{1}du \left(
\frac{1}{e^{B_0u+\frac{\mu}{k_BT}}+1}+\frac{1}{e^{B_0u-\frac{\mu}{k_BT}}+1}\right)
\sqrt{1-u^2}.
\nonumber
\end{eqnarray}

Owing to the condition (\ref{eq18}), the first contribution on the right-hand side
of (\ref{eq22}) is much less than the second and can be neglected. Owing to the same
condition, according to (\ref{eq15}),

\begin{equation}
B_0=0=\frac{\hbar c\Fv y}{4ak_BT}\approx\frac{\Fv\hbar\omega_c}{2k_BT}\ll 1.
\label{eq23}
\end{equation}

Because of this, it holds $B_0u\ll 1$ and one can put $\exp(B_0u)\approx 1$ in
(\ref{eq22}). As a result, (\ref{eq22}) takes the form

\begin{eqnarray}
&&
\pP\approx\frac{8\alpha k_BT}{\Sv\hbar \omega_c}
\ln\left[\left(1+e^{\frac{\mu}{k_BT}}\right)
\left(1+e^{-\frac{\mu}{k_BT}}\right)\right]
\nonumber\\[0mm]
&&\label{eq24}\\[-2mm]
&&-\frac{4\alpha}{\Fv} \left[\left(
1+e^{\frac{\mu}{k_BT}}\right)^{-1}+\left(1+e^{-\frac{\mu}{k_BT}}\right)^{-1}\right]
\int\limits_{0}^{1}du\sqrt{1-u^2}.
\nonumber
\end{eqnarray}

Calculating the integral, we find an expression

\begin{eqnarray}
&&
\pP\approx\frac{\alpha}{\Fv}\left\{\frac{8 k_BT}{\Fv\hbar \omega_c}
\ln\left[\left(1+e^{\frac{\mu}{k_BT}}\right)
\left(1+e^{-\frac{\mu}{k_BT}}\right)\right]\right.
\nonumber\\[0mm]
&&\label{eq25}\\[-2mm]
&&-\left.\pi \left[\left(
1+e^{\frac{\mu}{k_BT}}\right)^{-1}+\left(1+e^{-\frac{\mu}{k_BT}}\right)^{-1}\right]
\right\},
\nonumber
\end{eqnarray}

\noindent
where, thanks to (\ref{eq18}), the second term is much less than the first.
As a result, one obtains

\begin{equation}
\pP\approx\frac{8\alpha k_BT}{\Sv\hbar \omega_c}
\ln\left[\left(1+e^{\frac{\mu}{k_BT}}\right)
\left(1+e^{-\frac{\mu}{k_BT}}\right)\right].
\label{eq26}
\end{equation}

In the special case of a pristine graphene $\Delta=\mu=0$,  (\ref{eq26}) reduces to

\begin{equation}
\pP\approx\frac{16\alpha k_BT}{\Sv\hbar \omega_c}
\ln 2,
\label{eq27}
\end{equation}

\noindent
which agrees with \cite{81}.

We are coming now to the case of arbitrary, but not too small, values of $\Delta$ and
any value of $\mu$. In fact we assume that $D_0$ defined in (\ref{eq15}) with $y=1$
is much larger than unity

\begin{equation}
D_0=\frac{2a\Delta}{\hbar c\Fv y}\approx \frac{\Delta}{\hbar \omega_c\Fv}\gg 1.
\label{eq28}
\end{equation}

\noindent
The assumption (\ref{eq28}) is not too restrictive. The point is that we consider
the Casimir-Polder force at large separations $a>2~\mu$m, i.e., $\hbar\omega_c<0.05~$eV.
This means that the condition (\ref{eq28}) is satisfied for all $\Delta>0.001~$eV.

Let us consider the first term in the polarization tensor (\ref{eq14}) with $y=1$.
Using the definition of $\Psi$ in (\ref{eq8}) and expanding arctan$(D_0^{-1})$ in powers of
small parameter $D_0^{-1}$, we obtain

\begin{equation}
\Psi(D_0)\approx\frac{8}{3D_0}, \qquad
\frac{\alpha}{\Fv}\Psi(D_0)\approx\frac{8\alpha}{3\Fv D_0}=
\frac{8\alpha}{3}\frac{\hbar\omega_c}{\Delta}.
\label{eq29}
\end{equation}

\noindent
The maximum value of the latter quantity in (\ref{eq29}) [i.e., of the first term in (\ref{eq14})]
for our values of parameters is unity and it decreases with increasing $\Delta$. Thus, thanks
to (\ref{eq18}), the first term in (\ref{eq14}) is much less that the second one containing the
logarithm function.

We turn our attention to the third term in (\ref{eq14}). Due to (\ref{eq28}), the lower and
upper integration limits are very close and one can replace $B_0u$ with $B_0D_0$ in the powers
of exponents entering $w_0(u,y)$ defined in (\ref{eq9}).
Taking into account that, according to (\ref{eq15}), $B_0D_0=\Delta/(2k_BT)$,
we can rewrite (\ref{eq14}) with $y=1$ in the form

\begin{eqnarray}
&&
\pP\approx\frac{8\alpha k_BT}{\Sv\hbar \omega_c}
\ln\left[\left(e^{-\frac{\Delta}{2k_BT}}+e^{\frac{\mu}{k_BT}}\right)
\left(e^{-\frac{\Delta}{2k_BT}}+e^{-\frac{\mu}{k_BT}}\right)\right]
\label{eq30}\\
&&-\frac{4\alpha}{\Fv} \left(
\frac{1}{e^{\frac{\Delta}{2k_BT}+\frac{\mu}{k_BT}}+1}+
\frac{1}{e^{\frac{\Delta}{2k_BT}-\frac{\mu}{k_BT}}+1}\right)
\int\limits_{D_0}^{\sqrt{1+D_0^2}}du\frac{1-u^2}{\sqrt{1-u^2+D_0^2}}.
\nonumber
\end{eqnarray}

\noindent
The integral in (\ref{eq30}) is easily calculated

\begin{equation}
I=\int\limits_{D_0}^{\sqrt{1+D_0^2}}du\frac{1-u^2}{\sqrt{1-u^2+D_0^2}}=
-\frac{D_0}{2}+\frac{D_0^2-1}{2}\left({\rm arctan}D_0-\frac{\pi}{2}\right).
\label{eq31}
\end{equation}

Under the condition (\ref{eq28}), we find from (\ref{eq31}) $I\approx -D_0$ and
(\ref{eq30}) leads to

\begin{eqnarray}
&&
\pP\approx\frac{8\alpha k_BT}{\Sv\hbar \omega_c}
\ln\left[\left(e^{-\frac{\Delta}{2k_BT}}+e^{\frac{\mu}{k_BT}}\right)
\left(e^{-\frac{\Delta}{2k_BT}}+e^{-\frac{\mu}{k_BT}}\right)\right]
\nonumber\\
&&+\frac{4\alpha\Delta}{\Sv\hbar\omega_c} \left(
\frac{1}{e^{\frac{\Delta}{2k_BT}+\frac{\mu}{k_BT}}+1}+
\frac{1}{e^{\frac{\Delta}{2k_BT}-\frac{\mu}{k_BT}}+1}\right)
.
\label{eq32}
\end{eqnarray}

\noindent
After making identical transformations  in the first and second terms of this expression,
we bring it to the form [see (\ref{A7}) in Appendix A for details]

\begin{eqnarray}
&&
\pP\approx\frac{8\alpha k_BT}{\Sv\hbar \omega_c}\left[
\ln\left(4\cosh\frac{\Delta+2\mu}{4k_BT}\cosh\frac{\Delta-2\mu}{4k_BT}\right)
\right.
\nonumber\\
&&
\left.-\frac{\Delta}{4k_BT} \left(
\tanh\frac{\Delta+2\mu}{4k_BT}+\tanh\frac{\Delta-2\mu}{4k_BT}
\right)\right]
.
\label{eq33}
\end{eqnarray}

By putting $\mu=0$ in (\ref{eq33}), one finds

\begin{equation}
\pP\approx\frac{16\alpha k_BT}{\Sv\hbar \omega_c}\left[
\ln\left(2\cosh\frac{\Delta}{4k_BT}\right)
-\frac{\Delta}{4k_BT}
\tanh\frac{\Delta}{4k_BT}\right]
.
\label{eq34}
\end{equation}

\noindent
Under the additional condition $\Delta\ll 4k_BT$, we can neglect by the second term in
(\ref{eq34}), as compared to the first one, and obtain

\begin{equation}
\pP\approx\frac{16\alpha k_BT}{\Sv\hbar \omega_c}
\ln\left(2\cosh\frac{\Delta}{4k_BT}\right)
.
\label{eq35}
\end{equation}

\noindent
This result coincides with that obtained earlier in \cite{81} if to take into account
that \cite{81} uses the notation $\tilde{\Delta}=\Delta/(\hbar\omega_c)$, where $\Delta$
is equal to $\Delta/2$ in our current notations, i.e., to one half of the total energy gap.

Note that at $T=300~$K the application region of (\ref{eq35}) reduces to
$0.001~\mbox{eV}<\Delta<0.01~$eV, i.e., it is rather narrow. In the Appendix A, using the
condition opposite to (\ref{eq28}), we prove, however, that (\ref{eq35}) remains valid
for arbitrary small values of $\Delta$ [see Equation (\ref{A8}) with any $\mu$ including
$\mu=0$].

Now we finalize the asymptotic expression $F_0^{as}$ for the Casimir-Polder force from
gapped and doped graphene with not too small energy gap $\Delta$. For this purpose, we
substitute (\ref{eq33}) to (\ref{eq21}). The obtained expression  $F_0^{as}$ is valid
under the condition (\ref{eq18}). In the next section, we find how close would the
asymptotic Casimir-Polder force be to the numerical values of the force at large
separations $F_0$.

\section{Comparison Between Asymptotic and Numerical Results}

Here, we compare the analytic asymptotic expressions for the large-separation
Casimir-Polder force $F_0^{as}$ obtained in Section 4 with numerical computations
of $F_0$ for
different values of the energy gap and chemical potential.

We begin with the case of an undoped graphene sheet, $\mu =0$, and calculate the
ratio $F_0/F_0^{as}$ for different values of the energy gap $\Delta$. In doing so,
$F_0$ is computed by (\ref{eq12})--(\ref{eq15}) and $F_0^{as}$ by (\ref{eq21})
and (\ref{eq34}). All computations are performed at $T=300\,$K.

In Figure~\ref{fg3}, the ratio $F_0/F_0^{as}$ is shown as a function of separation
between an atom (nanoparticle) and a graphene sheet by the three lines counted
from top to bottom for the energy gap $\Delta$=0.1, 0.15, 0.2\,eV, respectively.
The case of large separations up to 100\,$\mu$m is shown in the inset for
$\Delta$=0.15 and 0.2\,eV.

\begin{figure}[H]
\vspace*{-5.2cm}
\centerline{\hspace*{-2.7cm}
\includegraphics[width=5in]{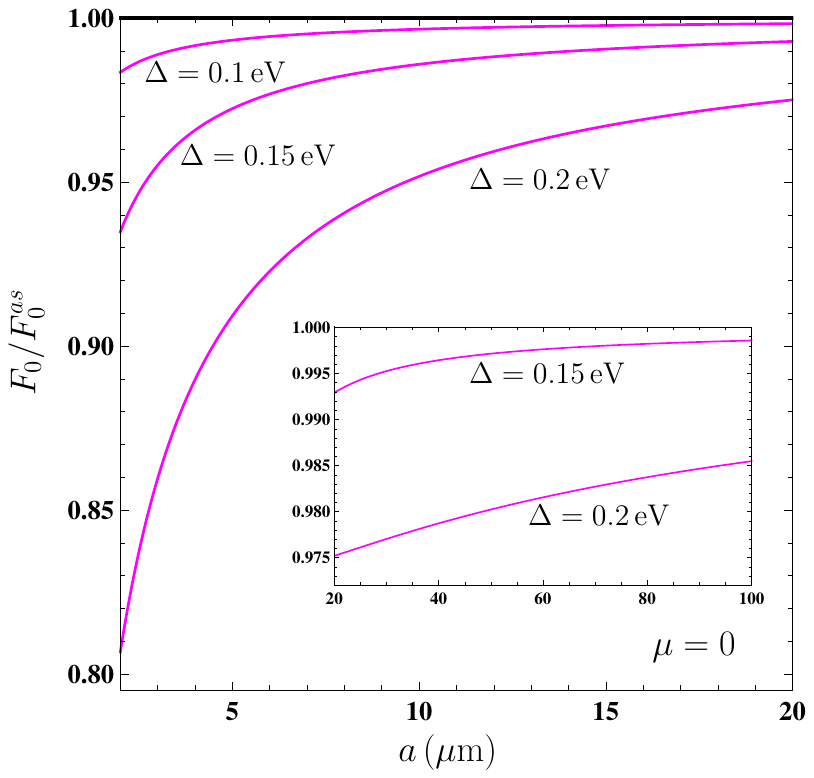}}
\vspace*{-4.7cm}
\caption{\label{fg3}
The ratio of the Casimir-Polder force from a graphene sheet with $\mu = 0$
at large separations to its asymptotic behavior is shown as a function of separation by the
three lines counted from top to bottom for the energy gap $\Delta$ equal to 0.1, 0.15, and
0.2\,eV. In the inset, the two lines for $\Delta = 0.15$ and 0.2\,eV are shown at larger
separations.}
\end{figure}

As is seen in Figure~\ref{fg3}, the best agreement between the asymptotic and
computed Casimir-Polder forces holds for the smallest $\Delta$=0.1\,eV. In this case,
$F_0^{as}$ agrees with $F_0$ in the limits of 1\% at any $a > 3\,\mu$m. With
increasing $\Delta$, an agreement between $F_0^{as}$ and $F_0$ gets worse.
Thus, for a graphene sheet with $\Delta$=0.15\,eV the 1\% agreement is reached
at $a$=14\,$\mu$m. As to graphene with $\Delta$=0.2\,eV, the 2\% agreement
is reached only at  $a$ = 50\,$\mu$m.

Now we consider an impact of the chemical potential on the measure of agreement
between $F_0^{as}$ and $F_0$. For this purpose, we consider the graphene sheets
with $\Delta$=0.2\,eV (the case of the worst agreement in Figure~\ref{fg3}) but
various values of the chemical potential. Computations of $F_0^{as}$ are
performed by (\ref{eq21}) and (\ref{eq33}).

In Figure~\ref{fg4}, the ratio $F_0/F_0^{as}$ is again shown as a function of separation
by the three lines counted from top to bottom for the chemical potential $\mu = 150$, 75,
and 25\,meV, respectively (brown, blue, and red lines). In the inset, the lines for a graphene
sheet with $\mu = 75$ and 25\,meV are shown in the region of large separations up to
100\,$\mu$m.

{}From Figure~\ref{fg4}, one can conclude that an increase in the value of the chemical
potential makes an agreement between $F_0^{as}$ and $F_0$ better. Thus, for
$\mu = 150$\,meV the 1\% agreement occurs at all separations $a > 3$\,$\mu$m,
whereas for $\mu = 75$\,meV at $a > 5.5$\,$\mu$m. For a graphene sheet with
$\mu = 25$\,meV the 1\% agreement is reached only at  $a \approx 34$\,$\mu$m.
We can say that an increase in the values of $\Delta$ and $\mu$ acts on an agreement
between $F_0^{as}$ and $F_0$ in the opposite directions by making it worse and
better, respectively, at the same separation distance.

The above results allow to determine the region of distances where the large-separation
Casimr-Polder force $F_0$ can be replaced with its asymptotic behavior $F_0^{as}$
depending on the values of the energy gap and chemical potential of the specific
graphene sheet. These results are valid for both light and heavy atoms and for
spherical nanoparticles.

\begin{figure}[H]
\vspace*{-5.2cm}
\centerline{\hspace*{-2.7cm}
\includegraphics[width=5in]{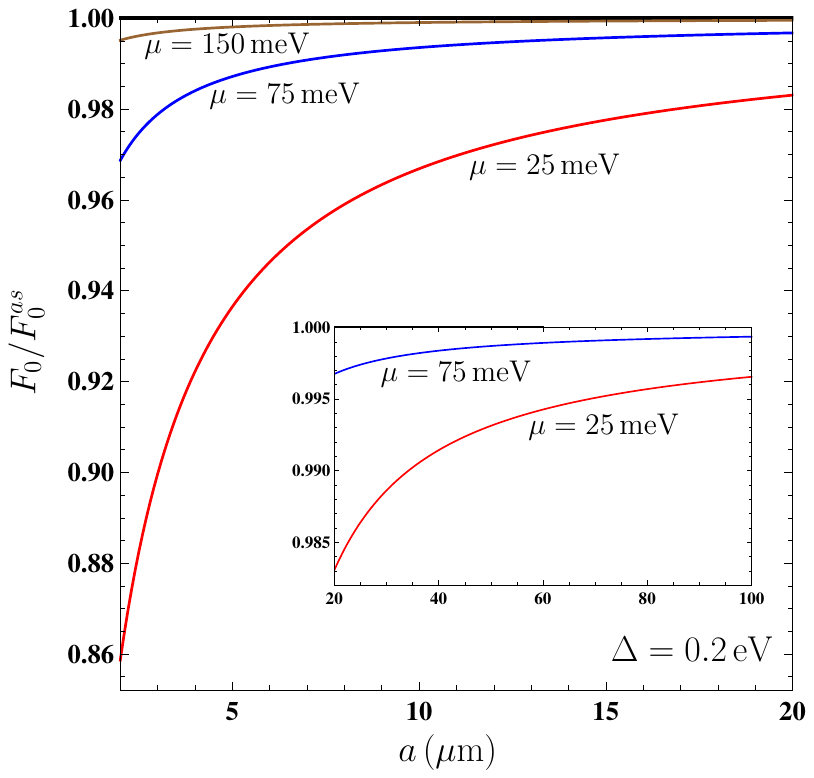}}
\vspace*{-4.7cm}
\caption{\label{fg4}
The ratio of the Casimir-Polder force from a graphene sheet with $\Delta = 0.2$\,eV
at large separations to its asymptotic behavior is shown as a function of separation by the
three lines counted from top to bottom for the chemical potential $\mu$ equal to 150, 75, and
25\,meV. In the inset, the two lines for $\mu = 75$ and 25\,meV are shown at larger
separations. }
\end{figure}

\section{Discussion}
As discussed in Section 1, the Casimir-Polder force on atoms and nanoparticles from
different surfaces including graphene is the subject of topical investigations in the
interests of both fundamental physics and its applications. The Casimir-Polder force from
graphene attracts an especial attention because graphene is the novel material of high
promise due to its unusual mechanical and electrical properties.

From the theoretical point of view, graphene offers major advantages over the more
conventional materials because its response functions to the electromagnetic field
can be found on the basis of first principles of thermal quantum field theory without
resort to phenomenological models. This is not the case for real metals whose response
to the low-frequency electromagnetic field is described by the phenomenological
Drude model, which lacks an experimental confirmation in the area of s-polarized
evanescent waves giving an important contribution to the Casimir effect \cite{82,83}.
As a result, there are contradictions between the predictions of the Lifshitz theory and
measurements of the Casimir force between metallic surfaces (see \cite{77,80,84,85}
for a review).

Although the Casimir-Polder force from graphene is not yet measured, the already
performed measurements of the Casimir force between a graphene-coated plate and
an Au-coated sphere demonstrate an excellent agreement between theoretical
predictions of the Lifshitz theory using the polarization tensor of graphene and the
measurement data \cite{86,87}. Because of this, the above results for the Casimir-Polder
force from gapped and doped graphene at large separations, obtained here using the
formalism of the polarization tensor, are of high degree of reliability.

\section{Conclusions}

To conclude, in the foregoing we investigated the Casimir-Polder force acting on atoms
and nanoparticles from the gapped and doped graphene sheet at large separations.
We have found separation distances starting from which the zero-frequency term of the
Lifshitz formula coincides with the total Casimir-Polder force acting on heavy atoms
or spherical nanoparticles in the limits of 1\%.
It was shown that, depending on the values of the energy gap and chemical potential
of graphene, the classical limit may be reached at much larger distances than the
limit of large separations.

Furthermore, we derived the analytic asymptotic expressions for the zero-frequency
term of the Lifshitz formula at large separations with the reflection coefficient
expressed via the polarization tensor of graphene. These expressions are valid
for light and heavy atoms and nanoparticles of spherical shape. The obtained asymptotic
expressions were compared with numerical computations of the zero-frequency term.
According to our results, with increasing energy gap of graphene, the separation
distance ensuring a better than 1\% agreement between the asymptotic and numerically
computed forces also increases. By contrast, an increase of the chemical potential
of graphene leads to a 1\% agreement between the asymptotic and numerical
results at shorter separations.

The obtained results make it possible to easily calculate the large-separation
Casimir-Polder force from the gapped and doped graphene sheets and to control
it by varying the values of the energy gap and chemical potential. This can be used
in precision experiments on quantum reflection and Bose-Einstein condensation
near the surfaces of graphene, as well as in various technological applications.
In future it would be interesting to investigate the large-separation Casimir-Polder
force from the graphene-coated substrates made of different materials.

\vspace{6pt}

\funding{G.L.K. was partially funded by the
Ministry of Science and Higher Education of Russian Federation
("The World-Class Research Center: Advanced Digital Technologies,"
contract No. 075-15-2022-311 dated April 20, 2022). The research
of V.M.M. was partially carried out in accordance with the Strategic
Academic Leadership Program "Priority 2030" of the Kazan Federal
University. }

\appendixtitles{yes}
\appendixstart
\appendix
\section{Asymptotic Expression for Graphene with Small Energy Gap }
\setcounter{equation}{0}
\renewcommand{\theequation}{A\arabic{equation}}

The asymptotic expression for the Casimir-Polder force from gapped and doped graphene
obtained in Section~4 is valid for graphene, satisfying the condition (\ref{eq28}), i.e.,
having not too small energy gap. Now we consider the separation region where the
condition (\ref{eq18}) is again satisfied but the energy gap satisfies the condition

\begin{equation}
D_0\approx\frac{\Delta}{\Fv\hbar\omega_c}\ll 1,
\label{A1}
\end{equation}

\noindent
which is just the opposite to (\ref{eq28}).

Owing the condition (\ref{eq18}), the inequality (\ref{eq23}) preserves its validity and
the first contribution to the polarization tensor (\ref{eq14}) with $y=1$ is much less than
the second and can be omitted.

First, we evaluate the third contribution to (\ref{eq14}) given by

\begin{equation}
I(1)=-\frac{4\alpha}{\Fv}\int\limits_{D_0}^{\sqrt{1+D_0^2}}du
\left(
\frac{1}{e^{B_0u+\frac{\mu}{k_BT}}+1}+\frac{1}{e^{B_0u-\frac{\mu}{k_BT}}+1}\right)
\frac{1-u^2}{\sqrt{1-u^2+D_0^2}}.
\label{A2}
\end{equation}

\noindent
By introducing the new integration variable, $v=u-D_0$, this term takes the form

\begin{eqnarray}
&&
I(1)=-\frac{4\alpha}{\Fv}\int\limits_{0}^{\sqrt{1+D_0^2}-D_0}dv
\left(
\frac{1}{e^{\frac{\Delta}{2k_BT}+\frac{\mu}{k_BT}}e^{B_0v}+1}+
\frac{1}{e^{\frac{\Delta}{2k_BT}-\frac{\mu}{k_BT}}e^{B_0v}+1}\right)
\nonumber \\
&&~~~~\times
\left(\sqrt{1-v^2-2vD_0}-\frac{D_0^2}{\sqrt{1-v^2-2vD_0}}\right).
\label{A3}
\end{eqnarray}

\noindent
Using  (\ref{A1}), we conclude that the upper integration limit in
(\ref{A3}) $\sqrt{1+D_0^2}-D_0\sim 1$. Then, because of (\ref{eq23}), one can put
$\exp(B_0v)\approx 1$ and rewrite (\ref{A3}) as

\begin{eqnarray}
&&
I(1)\approx -\frac{4\alpha }{\Fv}
\left(
\frac{1}{e^{\frac{\Delta}{2k_BT}+\frac{\mu}{k_BT}}+1}+
\frac{1}{e^{\frac{\Delta}{2k_BT}-\frac{\mu}{k_BT}}+1}\right)
\nonumber \\
&&~~~~\times
\int\limits_{0}^{\sqrt{1+D_0^2}-D_0}dv\left(\sqrt{1-v^2-2vD_0}-\frac{D_0^2}{\sqrt{1-v^2-2vD_0}}\right).
\label{A4}
\end{eqnarray}

Calculating the integral in (\ref{A4}), we obtain

\begin{eqnarray}
&&
I(1)\approx -\frac{2\alpha}{\Fv}
\left(
\frac{1}{e^{\frac{\Delta}{2k_BT}+\frac{\mu}{k_BT}}+1}+
\frac{1}{e^{\frac{\Delta}{2k_BT}-\frac{\mu}{k_BT}}+1}\right)
\nonumber \\
&&~~~~\times
\left[(1-D_0^2)\frac{\pi}{2}-D_0-(1+D_0^2){\rm arctan}D_0\right]
\label{A5}\\
&&
=-\frac{2\alpha}{\Fv}
\left(
\frac{1}{e^{\frac{\Delta}{2k_BT}+\frac{\mu}{k_BT}}+1}+
\frac{1}{e^{\frac{\Delta}{2k_BT}-\frac{\mu}{k_BT}}+1}\right)
\left[\pi+O(D_0)\right].
\nonumber
\end{eqnarray}

\noindent
With account of (\ref{eq18}), it is seen that the magnitude of $I(1)$ is much
less than the second term in the polarization tensor (\ref{eq14}) and can be
omitted.

Thus, we are left with only the second term in (\ref{eq14})

\begin{equation}
\pP\approx\frac{8\alpha k_BT}{\Sv\hbar \omega_c}
\ln\left[\left(e^{-\frac{\Delta}{2k_BT}}+e^{\frac{\mu}{k_BT}}\right)
\left(e^{-\frac{\Delta}{2k_BT}}+e^{-\frac{\mu}{k_BT}}\right)\right],
\label{A6}
\end{equation}

\noindent
which can be transformed similar to (\ref{eq32}). Here, we present this transformation
in greater detail

\begin{eqnarray}
&&
\pP\approx\frac{8\alpha k_BT}{\Sv\hbar \omega_c}
\ln\left[e^{\frac{\mu}{k_BT}}\left(e^{-\frac{\Delta}{2k_BT}-\frac{\mu}{k_BT}}+1\right)
e^{-\frac{\mu}{k_BT}}\left(e^{-\frac{\Delta}{2k_BT}+\frac{\mu}{k_BT}}+1\right)\right]
\nonumber \\
&&
=\frac{8\alpha k_BT}{\Sv\hbar \omega_c}
\ln\left[e^{-\frac{\Delta+2\mu}{4k_BT}}\left(
e^{\frac{\Delta+2\mu}{4k_BT}}+e^{-\frac{\Delta+2\mu}{4k_BT}}\right)
e^{-\frac{\Delta-2\mu}{4k_BT}}\left(
e^{\frac{\Delta-2\mu}{4k_BT}}+e^{-\frac{\Delta-2\mu}{4k_BT}}\right)\right]
\nonumber \\
&&
=\frac{8\alpha k_BT}{\Sv\hbar \omega_c}\ln\left(4\cosh\frac{\Delta+2\mu}{4k_BT}
\cosh\frac{\Delta-2\mu}{4k_BT}\right)-\frac{2\alpha\Delta}{\Sv\hbar\omega_c}.
\label{A7}
\end{eqnarray}

Owing to condition (\ref{A1}), the last term in (\ref{A7}) can be neglected and,
as a result,

\begin{equation}
\pP\approx\frac{8\alpha k_BT}{\Sv\hbar \omega_c}
\ln\left(4\cosh\frac{\Delta+2\mu}{4k_BT}\cosh\frac{\Delta-2\mu}{4k_BT}\right).
\label{A8}
\end{equation}

For $\mu=0$, (\ref{A8}) reduces to (\ref{eq35}). Thus, (\ref{eq35}) is really valid
for arbitrary small $\Delta$ satisfying the condition (\ref{A1}).

\begin{adjustwidth}{-\extralength}{0cm}

\reftitle{References}

\end{adjustwidth}
\end{document}